%% file: paper.tex
\documentclass[aps,twocolumn]{revtex4-2}
\usepackage{amsmath,mathtools}
\usepackage{amssymb}

\usepackage[export]{adjustbox}

\usepackage{xcolor}
\usepackage{graphicx}

\newcommand{\beq}{\begin{equation}} 
\newcommand{\eeq}{\end{equation}} 
  
\newcommand{\bev}{\begin{verbatim}} 
\newcommand{\eev}{\end{verbatim}}

\newcommand{\frc}[2]{\mbox{$\frac{#1}{#2}$}}

\newcommand{\epa}{\mathcal{E}} 
\newcommand{\tpa}{\mathcal{T}}  
\newcommand{\ceff}{C_{\mathrm{eff}}} 
\newcommand{\deff}{D_{\mathrm{eff}}} 
\newcommand{\EDF}{\mathcal{H}} 
\newcommand{\imaxx}{i_{\mathrm{max}}} 

\newcommand{\wiv}{W_1} 
\newcommand{\Ksym}{K_{\mathrm{sym}}} 
\newcommand{\Qsym}{Q_{\mathrm{sym}}} 

\begin{document}

\title{CREX and PREX-II Reconciled within Energy-Density Functional Theory} 

\author{Panagiota Papakonstantinou}

\affiliation{Institute for Rare Isotope Science, 
                        Institute for Basic Science, Daejeon 34000, Republic of Korea; ppapakon@ibs.re.kr}

\begin{abstract} 
The CREX and PREX-II measurements of the neutron-skin thickness of 
$^{48}$Ca and $^{208}$Pb challenge standard nuclear energy-density functional (EDF) descriptions of nuclei and nuclear matter. 
It is shown that the apparent tension arises from an implicit constraint in current EDF parameterizations that ties the density dependence of the functional at the dilute nuclear surface to that of uniform matter near saturation.
Relaxing this surface-bulk coupling and independently constraining the dilute-density sector,
while preserving realistic saturation properties and the high-density equation of state, yields EDFs that simultaneously reproduce the neutron skins of both nuclei, their electric dipole polarizabilities, and neutron-star mass–radius relations. 
The established correlation between the neutron-skin thickness and the symmetry-energy slope parameter $L$ at saturation is retained but becomes substantially broader. The results show that current neutron-skin data do not require extreme values of $L$ and highlight an underconstrained degree of freedom associated with low-density nuclear matter.
\end{abstract}
\maketitle

\section{Introduction}

Standard energy-density functional (EDF) studies reveal a strong correlation between the neutron-skin thickness $R_{np}$ of nuclei with neutron excess 
and the stiffness of the nuclear equation of state (EoS).
EDF models with a large symmetry-energy slope parameter $L$ (stiff EoSs) predict thicker neutron skins, while models with a small $L$ (soft EoSs) predict thinner ones~\cite{RoP2018}. 
This correlation has been challenged by the CREX measurement of the thin neutron skin of $^{48}$Ca~\cite{CREX}, favoring soft models, and the PREX-II measurement of the thick neutron skin of $^{208}$Pb~\cite{PREXII}, favoring stiff models.  
A related strong tension exists between the moderate dipole polarizability of $^{208}$Pb, 
favoring softer EoSs, and its thick neutron skin~\cite{Pie2021}. 
Understanding these tensions is essential for nuclear structure and multimessenger astrophysics~\cite{BFH2021,MaF2024}. 

Ensuing efforts to reconcile the CREX and PREX-II measurements, including working directly with the measured parity-violating asymmetry and revisiting the data analyses, have not succeeded~\cite{RRN2022,YuP2023,RoJ2025,ReH202X}.  
A Bayesian analysis in the meta-model affirmed the tensions~\cite{MoG2023}. 
More recently, it was revealed that the isovector spin-orbit force plays a role in predictions for $R_{np}$, but reconciliation was found possible only for unphysical values that are incompatible with shell structure~\cite{Zha2025,KPS2025} unless tensor terms were included~\cite{YZC2026}. 
Alpha clustering, too, was found to affect the $R_{np}$ of the two nuclei to different degrees, setting the relationship between $L$ and $R_{np}$ up for revision~\cite{YLX2023}. 
The role of alpha clustering in $R_{np}$ predictions had been examined earlier in Sn and Pb isotopes \cite{Typ2014}. 
A reconciliation of the CREX and PREX-II measurements through an enhanced isovector tensor coupling in a relativistic framework was reported very recently~\cite{QYZ202X}.
However, reconciliation of those with neutron-star properties and with dipole polarizabilities was not reported. 
Similarly, an exploration of the curvature parameter of the symmetry energy, $K_{\mathrm{sym}}$, has demonstrated reconciliation of the  PREX-II result  
and the neutron-star tidal deformability only~\cite{GuN2025}. 

Free explorations of $K_{\mathrm{sym}}$ are very important for addressing artificial model correlations between EoS parameters and specific nuclear properties~\cite{NeC2021,XuP2022} but do not suffice for reconciling the two $R_{np}$ values. 
Only free and extreme variations of higher-order parameters at the same time, namely the skewness and kurtosis, have succeeded in reconciliation~\cite{rila2022}. 
Such an approach is not viable because it affects both the dilute and dense matter EoS on an equal footing. 
However, its tentative success points to a resolution through directly addressing the EoS of dilute matter, which thus far remains largely unconstrained~\cite{MoG2023,HoS2006}. 

In the present work, a reconciliation of 
the neutron-skin thickness measurements, the dipole polarizabilities, and the EoS of dense matter is achieved by directly exploring the dilute sector. 
The premise is that the EoS of the dilute matter that forms on the nuclear surface need not be a straightforward extrapolation from the saturation regime. 
At low densities, nucleonic matter may be in a gas state with or without clusters rather than a uniform quantum liquid. 
Its description on the nuclear surface involves the additional complexity of shell structure. 
Consistent with the philosophy of EDF theory, all many-body correlations are absorbed into the functional regardless of their microscopic origin. 
A targeted modification is therefore introduced to the EDF to decouple the two regimes. 


\section{Method} 
The present strategy is to keep the bulk equation of state unchanged while introducing the minimal additional freedom required to explore the dilute-density regime.

The energy per nucleon of 
uniform nucleonic matter at zero temperature is considered as a function of the nucleon density $\rho$ and isospin asymmetry $\delta$, $E({\rho,\delta})$. 
The expansion of $E(\rho,0)$ in terms of $\rho$ provides the EoS parameters characterizing isospin-symmetric matter around its saturation density $\rho_0$, 
\begin{equation}
E(\rho,0) = E_0 + \frac{1}{2} K_0 x^2 + \frac{1}{6}Q_0 x^3 + \ldots ,
\label{eq:Erho} 
\end{equation} 
where $x\equiv (\rho -\rho_0)/3\rho_0$. The expansion of $E(\rho,\delta)$ in powers of the asymmetry $\delta$ defines the symmetry energy 
$S(\rho) = \frac{1}{2} \left. \frac{\partial^2 E(\rho,\delta)}{\partial \delta^2} \right|_{\delta=0} $, 
whose expansion coefficients 
\begin{equation} 
 S(\rho) = J + Lx + \frac{1}{2} K_{\mathrm{sym}} x^2 + \frac{1}{6}Q_{\mathrm{sym}} x^3 + \ldots 
\label{eq:Srho} 
\end{equation}
characterize the EoS of asymmetric nuclear matter around the saturation density. 

In order to connect it with finite nuclei, the EoS is translated into a non-relativistic EDF for nuclear structure calculations within the KIDS framework, which is  reviewed in detail \cite{PaH2023}. 
The effective interaction is an auxiliary construct that is used to generate the EDF and has the form of an extended Skyrme interaction~\cite{BHR2003}  (see App.~\ref{Sec:kids}). The form of the density dependence is informed by an effective theory of many-body Fermion systems~\cite{kidsnm}.  
Unlike with standard EDFs, effective masses can be explored freely and independently from the bulk EoS~\cite{kids_nuclei1}.  The  flexibility of the KIDS framework has allowed detailed studies of the correlations between EoS parameters and nuclear and astrophysical observables~\cite{XuP2022,ZXP2023}. 

The starting point is a standard KIDS  EDF, fully determined via the parameters $(E_0,\rho_0,K_0)$ for the symmetric-matter EoS  
and $(J,L,K_{\mathrm{sym}},Q_{\mathrm{sym}})$ for the symmetry energy $S(\rho)$, 
as well as the isoscalar effective mass $(m^{\star}/m)$, isovector enhancement factor $\kappa$, isoscalar and isovector gradient terms $C_{12},D_{12}$, and the spin-orbit couplings $W_0$ (isoscalar) and $\wiv$ (isovector). 
Note that in previous works within the KIDS framework, $\wiv$ was set equal to $W_0$.   
The algebraic expressions through which a given set of EoS parameters and $(m^{\star}/m),\kappa,C_{12},D_{12}$ 
are translated into an effective interaction and functional for nuclear structure calculations \cite{PaH2023} are reproduced in App.~\ref{Sec:kids}.

To introduce additional flexibility in the dilute-density regime, the effective interaction is augmented with a term that is active only at low densities. 
The form chosen is a minimal effective Skyrme-like term~\cite{BHR2003} scaled by an exponential factor, 
\begin{equation}
 V_d=\frac{1}{6} (t_d + y_dP_{\sigma})\rho^{a_d}e^{-b_d\rho^2} \delta (\vec{r}_1-\vec{r}_2) . 
\label{eq:Vd} 
\end{equation}
\textcolor{black}{A comprehensive treatment would interpolate between two energy-density functional ansätze: one appropriated for saturated matter and one for the dilute regime. While desirable and physically well motivated, the additional parameters such an approach would introduce are not warranted at present, given the scarcity and large uncertainties of the available data.} 

Although simple, expression (\ref{eq:Vd}) can accommodate the most relevant behaviors for the EoS of dilute matter: a sharp or slow rise depending on $a_d$, a slow or fast fall-off depending on $b_d$, and an enhancement or suppression depending on the signs of the prefactors.  
It contributes to the symmetry energy $S(\rho)$ a term equal to 
\begin{equation} 
S_d(\rho)=-\frac{1}{48}(t_d+2y_d)\rho^{a_d+1}e^{-b_d\rho^2}.  
\end{equation} 
See also App.~\ref{Sec:Vd}. 

As will be verified below, the exponential factor is so small that $V_d$ and $S_d$ vanish already at densities of order $\rho_0/10$ and thus have no bearing on the properties of dense objects, especially the neutron-star mass-radius relation (NSMR). 
Typical calculations of the NSMR employ standard uniform EoSs only above the core-crust transition density, roughly equal to $2\rho_0/3$, and use EoSs specifically appropriate for the crust at lower densities~\cite{Vin2021}. 

To ensure a realistic description of neutron stars, I begin from EoS parameter sets that are already known to reproduce dense matter. 
Because the present modification vanishes well below saturation density, the bulk EoS parameters
$(\rho_0,E_0,K_0,J,L,K_{\mathrm{sym}},Q_{\mathrm{sym}} )$ will not be altered at all. 
The sets selected are APR P4~\cite{kids_nuclei2}, with 
\[ 
 (J,L,K_{\mathrm{sym}},Q_{\mathrm{sym}}) = (32.8,49.2,-156.3,583.1)\, \mathrm{MeV},  
\] 
from a fit of the KIDS model to the Akmal-Pandharipande-Ravenhall equation of state of neutron matter~\cite{apr}, 
QMC P4~\cite{kids_nuclei2}, with 
\[ 
(J,L,K_{\mathrm{sym}},Q_{\mathrm{sym}}) = (34.5,60.5,-88.9,751.2)\, \mathrm{MeV}, 
\] 
from a fit of the KIDS model to one of the Quantum Monte Carlo pseudo-data sets for neutron matter reported in \cite{qmc}, 
and  
\[ 
 (J,L,K_{\mathrm{sym}},Q_{\mathrm{sym}}) = (32.5,65,-180,650)\, \mathrm{MeV}, 
\] 
which was selected to represent the stiffer side of the acceptable ranges proposed in \cite{prc103}. 
These will be refered to as APR, QMC, and EoS(65), respectively. 
The EoS of symmetric nuclear matter is the same in all three cases, with 
\[ 
\rho_0=0.16~\mathrm{fm}^{-3}, E_0=-16~\mathrm{MeV}, K_0=240~\mathrm{MeV}. 
\] 
To each EoS, a dilute-matter modification is introduced via $V_d$. 
Unless otherwise specified, the isoscalar effective mass and isovector enhancement factor, $(m^{\star}/m,\kappa)$, are fixed at $(0.70,0.40)$, while the isovector gradient and isoscalar spin-orbit terms, $(D_{12},W_0)$, are set to the standard values $(5,133)$~MeV~fm$^5$. 

The quality of the description of closed-shell nuclei is quantified through the average deviation per datum (ADPD) over $N$ data, 
\begin{equation} 
\mathrm{ADPD}(N) = \frac{1}{N} \sum_{i=1}^N\left|  
\frac{O_i^{\mathrm{calc}}-O_i^{\mathrm{exp}}}{O_i^{\mathrm{exp}}} 
\right| ,
\label{eq:adpd} 
\end{equation} 
where $O_i$ represents one of the $N$ observables. The superscripts `$\mathrm{exp}$' and `$\mathrm{calc}$'  represent the experimental and calculated values, respectively. 
At present, $N=19$ input data are used corresponding to the binding energies and charge radii of 
$^{16}$O, $^{40,48}$Ca, $^{90}$Zr, $^{120,132}$Sn, and $^{208}$Pb  and in addition the energies of $^{56,68,78}$Ni, $^{100}$Sn, and $^{218}$U. 
The energy data are from the AME2020 compilation~\cite{AME2020} and the radius data from \cite{Angeli2013}. 

The parameter space is explored efficiently using the Metropolis-Hastings Monte Carlo (MHMC) method. 
First, the parameters that are sampled are those characterizing $V_d$, namely $(t_d,y_d,a_d,b_d)$, with fixed $(m^{\star}/m,\kappa,D_{12},W_0)$. 
In addition, the isoscalar gradient coefficient, $C_{12}$, and the isovector spin-orbit coupling, $\wiv$, are sampled.  
\textcolor{black}{The gradient and spin-orbit terms of the functional also warrant exploration because they depend on gradients of the density profile. They are therefore active on the surface and can affect the neutron skin.}

The prior distributions are set to uniform ones within broad ranges and such that the total contribution of $V_d$ to the energy amounts to a few MeV at most. 
For $\wiv$, only moderately high values are allowed in order to avoid unrealistic single-particle spectra~\cite{KPS2025}. Specifically, the ranges considered are $t_d=-10^4$ to $10^6$~MeV~fm$^{3+3a_d}$,  
$y_d=-10^7$ to $10^7$~MeV~fm$^{3+3a_d}$, 
$a_d=0.4$ to $1.2$, $b_d=5\times 10^3$  to $5\times 10^4$~fm$^6$ , 
$C_{12}=-85$ to $-70$~MeV~fm$^5$, and 
$\wiv=0$ to $250$~MeV~fm$^5$. 

The data or pseudo-data used for driving the iterations are 
the $R_{np}$ of $^{208}$Pb from the PREX-II experiment~\cite{PREXII} 
\[ R_{np}(^{208}\mathrm{Pb}) = 0.283 \pm 0.071\, \mathrm{fm}, \]
the $R_{np}$ of $^{48}$Ca from the CREX experiment~\cite{CREX}  
\[ R_{np}(^{48}\mathrm{Ca}) = 0.121 \pm 0.050\, \mathrm{fm}  \] 
including the reported model uncertainties, 
the ADPD for the data listed above 
\[ \mathrm{ADPD}(19) = 0 \pm 1.5 \% , \] 
and the symmetry energy at $\rho_d\equiv 0.005$~fm$^{-3}$ 
\[ S(\rho_d)  = 5 \pm 5 \, \mathrm{MeV} . \]
Although the ADPD is non-negative by construction, its central value was set to zero to favor minimal deviations. 
In the loose constraint on the symmetry energy at $\rho_d$,  
it was taken into account that $S(\rho)$ of cold matter at $0.002-0.005\,\mathrm{fm}^{-3}$ extracted from heavy-ion experiments is positive and no higher than $10$~MeV~\cite{Nat2010,Wada2012,HNR2014}. \textcolor{black}{Other existing constraints from heavy-ion collisions refer to higher densities and are generally consistent with the present EoSs.}

\section{Results} 
\subsection{Neutron skins, polarizability, neutron stars} 
\begin{figure} 
\includegraphics[width=8cm]{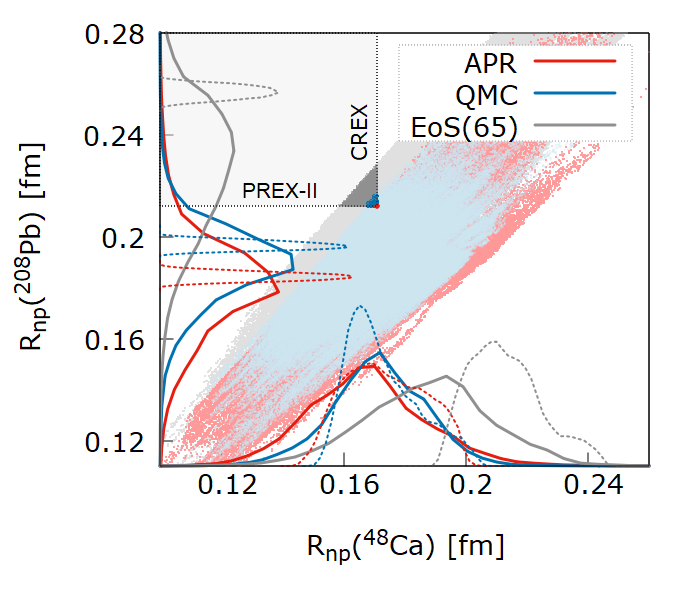} 
\caption{Neutron-skin thickness of $^{48}$Ca and $^{208}$Pb calculated with EDFs based on the shown bulk EoSs and satisfying ADPD$(19)\leq 1.5\%$. 
The rectangle encloses the reported experimental values. 
The curves correspond to posterior distributions (arbitrary common scale). 
{\em Dotted lines:} $V_d=0$ and varying $m^{\ast}/m,\kappa,W_0,D_{12}$; the results for $^{208}$Pb are divided by 5 for visibility.  
{\em Points and full lines:} Fixed $m^{\ast}/m,\kappa,W_0,D_{12}$ and varying $V_d$; points within the rectangle are highlighted.
 \label{fig:Rnp}} 
\end{figure}

Half a million MHMC iterations were performed for each EoS to sample the $(t_d,y_d,a_d,b_d,C_{12},\wiv)$ space.  
The hundreds of thousands of parameter sets satisfying ADPD$(19)\leq 1.5\%$ were then examined, thereby maintaining a good description of global nuclear properties.
In fig.~\ref{fig:Rnp}, the points and full lines visualize their distributions in comparison to the CREX and PREX-II results. 
The figure shows that a positive correlation between the mean $R_{np}$ and $L$ is preserved ({\em e.g.}, smaller $R_{np}$ for the softer APR), but even with all bulk EoS parameters fixed, including $L$, the $R_{np}$ values  span broad intervals. 
As a result, many EDFs are able to describe both measurements at the same time. 

For comparison, an analogous procedure was followed with $V_d$ set to zero: Instead of varying the four $V_d$ parameters, I varied  $(m^{\ast}/m,\kappa,W_0,D_{12})$ within the broad ranges $(0.5-1.0,0-1,100-200~\mathrm{MeV~fm}^5,-10~-~20~\mathrm{MeV~fm}^5)$, respectively. 
The resulting distributions for $R_{np}$ are shown with the dotted lines in Fig. ~\ref{fig:Rnp}.
They are very narrow for $^{208}$Pb and as a result, a simultaneous description of PREX-II and CREX is impossible. 
Clearly, the effect of $V_d$ cannot be reduced to the surface effects induced by the gradient, kinetic, and spin-orbit terms. 

Next, among the parameter sets including $V_d$ and satisfying ADPD$(19)\leq 1.5\%$, I consider as successful those for which, at the same time, 
$|R_{np}(^{208}\mathrm{Pb})-0.283| \leq 0.071 \, \mathrm{fm}$ 
and 
$|R_{np}(^{48}\mathrm{Ca})-0.121| \leq 0.050 \, \mathrm{fm} $, 
{\em i.e.}, for which each skin measurement is reproduced within the reported uncertainty.  
Under these strict conditions, 115 successful EDFs were found for QMC, and over ten thousand for the stiffer EoS(65). 
In this initial run, one successful EDF was identified for APR within the sampled parameter space, which will be denoted APR(A), but several others came close to satisfying all constraints. 
The abundance of solutions shows that reconciliation between PREX-II and CREX is possible for EoSs that are realistic in the saturation regime and beyond. 

The viability of each successful EDF will be assessed by means of the measured dipole polarizability $a_D$ of the two nuclei of interest, 
\[ 
    a_D(^{208}\mathrm{Pb}) = 20.1 \pm 0.6 \, \mathrm{fm}^3/e^2, 
\] 
as reported in \cite{Tam2011}, and 
\[ 
    a_D(^{48}\mathrm{Ca}) = 2.07 \pm 0.22  \, \mathrm{fm}^3/e^2,
\] 
as reported in \cite{Bir2017}. 
The dipole polarizability is calculated here within the self-consistent random-phase approximation (RPA) 
from the inverse energy-weighted sum rule $m_{-1}$
of the response function to the electric dipole operator, 
\[ a_D = \frac{8\pi e^2}{9} m_{-1} . \] 
The $a_D$ constraints were not included in the MHMC procedure because it takes substantially longer time to compute 
compared to the ground-state calculations with the Hartree-Fock code used in the MHMC iterations. 

Based on RPA runs with successful EDFs, the EDFs generated from the stiffer EoS(65) give $a_D$ values that are too high: Over $2.8$~fm$^3$ for $^{48}$Ca and over $24$~fm$^3$ for $^{208}$Pb. These EDFs, therefore, are deemed unviable. 
The successful sets based on the QMC EoS, on the other hand, give ranges that overlap with the data, 
yielding $a_D(^{208}\mathrm{Pb})$ up to $19.6$~fm$^3$ 
and $a_D(^{48}\mathrm{Ca}) = 2.21-2.32$~fm$^3$. 
One of those, which will be denoted QMC(A), describes both $a_D$ data   
within their reported uncertainties.
The APR(A) EDF gave the realistic $19.59$~fm$^3$ for $^{208}$Pb and the marginally realistic $2.31$~fm$^3$ for $^{48}$Ca.

\begin{figure} 
\includegraphics[width=9cm]{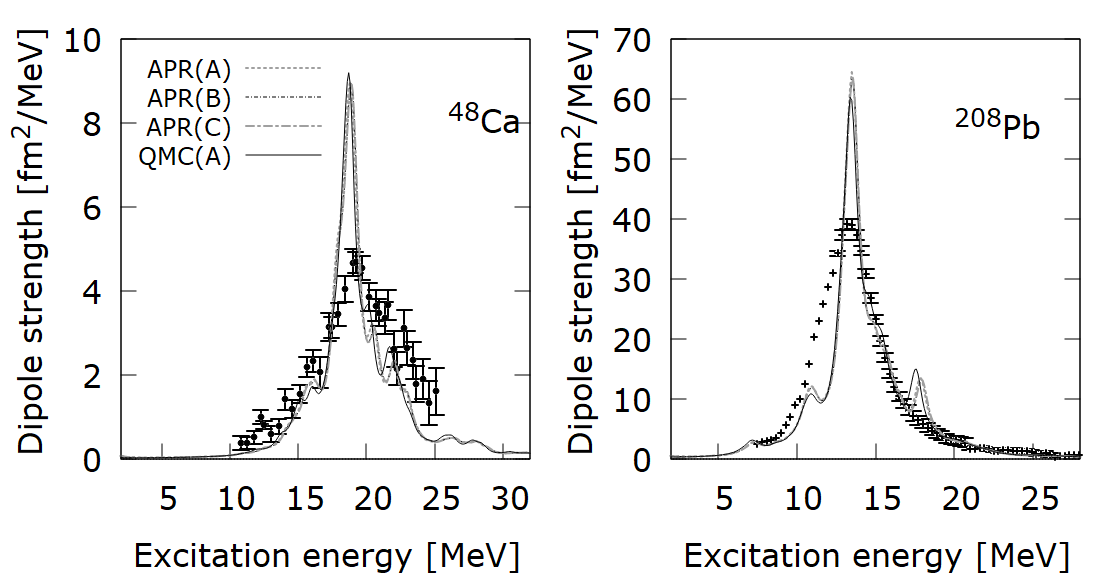}  
\caption{Dipole strength distributions obtained with APR(B), APR(C), and QMC(A), satisfying all four constraints, and APR(A), satisfying CREX, PREX-II, and $a_D(^{208}\mathrm{Pb})$.
Data are from the M0636 ($^{48}$Ca) and L0021 ($^{208}$Pb) data files of the EXFOR database. \label{fig:dipoles}} 
\end{figure} 

The parameter space for APR is worth exploring further. 
An additional iteration sequence of one and a half million steps was performed with no constraint on $S(\rho_d)$. 
The parameters were initialized to those of APR(A) and the step amplitudes were reduced. 
A total of $68$ successful (as defined earlier in terms of calculated ADPD(19) and neutron-skin thickness) EDF sets were obtained. 
With each of them, $a_D$ was calculated. 
All of them gave values for $a_D(^{208}\mathrm{Pb})$ between $19.5$ and $20.0$~fm$^3$, entirely within the experimental range. 
The improved result can plausibly be attributed to the removal of the constraint on $S(\rho_d)$, whose favored values for APR are found close to zero.  
Two of those also gave $a_D(^{48}\mathrm{Ca})$ values below $2.29$~fm$^3$ and are therefore consistent with all four observables. 
They will be denoted APR(B) and APR(C). 
Thus both APR and QMC EoSs succeed in describing all four nuclear observables. 
This result is noteworthy because the underlying bulk equations of state remain unchanged throughout.

Figure \ref{fig:dipoles} shows the dipole strength distribution calculated with APR(B), APR(C), and QMC(A), satisfying all four constraints, and APR(A), satisfying CREX, PREX-II, and $a_D(^{208}\mathrm{Pb})$.  
The differences are small and all calculations reproduce the photo-response well. 

For APR(B), APR(C), and QMC(A), and inserting $\rho_d$ as an indicative scaling, the values are 
$(t_d\rho_d^{a_d},y_d\rho_d^{a_d},a_d,b_d\rho_d^2)=
(4517,46436,0.875,0.795)$, 
$(4478,45947,0.877,0.795)$,
 and 
$(1177,17450,1.089,0.567)$, 
respectively, in units of 
$(\mathrm{MeV~fm}^{3},\mathrm{MeV~fm}^{3},1,1)$ 
and  $(C_{12},\wiv)=(-74.83,248.9), (-74.70,248.2), (-76.50,239.544)$~MeV~fm$^5$. 
The moderate enhancement observed in $\wiv$ compared to $W_0$ does not result in unphysical orbital ordering and energies. 
As an illustration, results with the APR(B) for the orbitals around the Fermi level in $^{208}$Pb are compared with empirical values in Table~\ref{table:levels}. 
\begin{table} 
\begin{tabular}{c|cc|| c | cc} 
 neutrons & exp & APR(B)  & protons & exp & APR(B) \\  
\hline 
$\nu 3p_{3/2}$   & $-8.25$   &  $-9.00$  & $\pi 1h_{11/2}$ &   $ -9.28$  & $-10.29$ \\ 
$\nu 2f_{5/2}$    &  $-7.88$   &  $-8.18$ & $\pi 2d_{3/2}$ &    $-8.32$    &  $-8.70$  \\ 
$\nu 3p_{1/2}$   & $-7.32$    &  $-7.54$ &  $\pi 3s_{1/2}$ &    $-7.95$    &  $-8.11$ \\
\hline
$\nu 2g_{9/2}$   & $-3.88$    &  $-3.65$ &$\pi 1h_{9/2}$ &    $-3.72$   &  $-3.07$ \\ 
\hline 
\end{tabular} 
\caption{Energies of single-particle levels in units of MeV calculated with the APR(B) EDF for $^{208}$Pb, compared with data compiled in \cite{SWV2007}. 
Three states below and one above the Fermi level are shown for neutrons and for protons. \label{table:levels} }  
\end{table} 
Finally, the neutron-skin thickness of $^{90}$Zr was calculated with the same EDFs and found between $0.072$ and $0.075$~fm, in agreement with values extracted from charge-exchange spin-dipole excitations~\cite{SYS2007,CWC2023}. 

\begin{figure} 
\includegraphics[width=7cm]{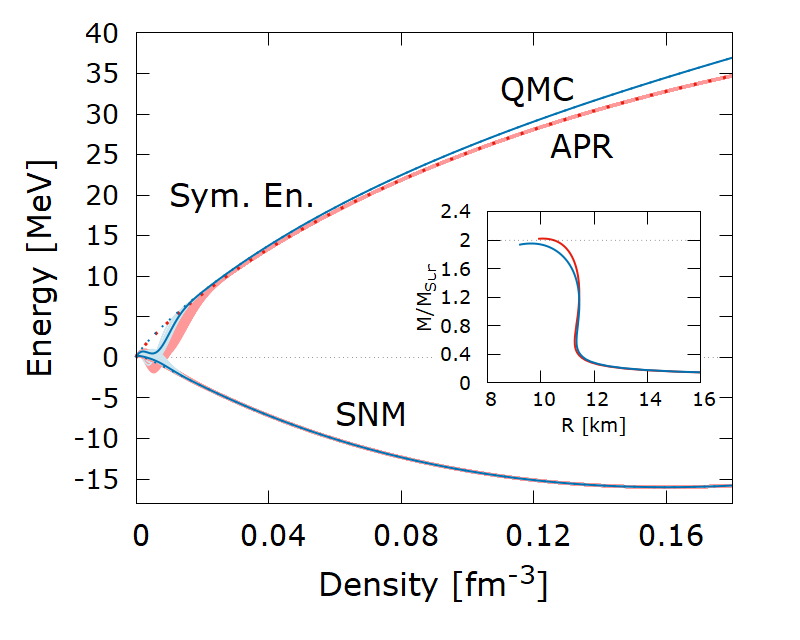}\\[-5mm]  
\caption{Energy per particle of symmetric nuclear matter (SNM) and symmetry energy for APR- and QMC-based EDFs (red and blue, respectively) before (dotted lines) and after (full lines) introducing the low-density modification. 
QMC(A) is highlighted with full blue lines. 
The inset shows the respective neutron star mass-radius relations. 
\label{fig:EoS}} 
\end{figure}

A large sample of the APR- and QMC-based EoSs that describe both the CREX and PREX-II data is shown in Fig.~\ref{fig:EoS}. 
The EoSs are also shown without the $V_d$ contribution. 
The respective NSMRs are shown to illustrate the high-density behavior. 
The symmetry energy displays a strong suppression below roughly $\rho_0/10$. 
This  seems to be a common feature in some preparatory explorations as well~\cite{rila2022,PapINPC}. 
In view of the experimental results pointing to a higher value of the symmetry energy at those densities \cite{Nat2010,Wada2012,HNR2014} and at the same time the ambiguity in interpreting the symmetry energy in a potentially clusterized regime and on the nuclear surface~\cite{Li2017,MoG2023}, one can view the present result as an {\em effective} symmetry energy absorbing diverse but genuine surface effects, which cannot be reduced to gradient terms such as $D_{12}$. 

\subsection{Other properties and discussion} 

\begin{figure} 
\includegraphics[width=8.5cm]{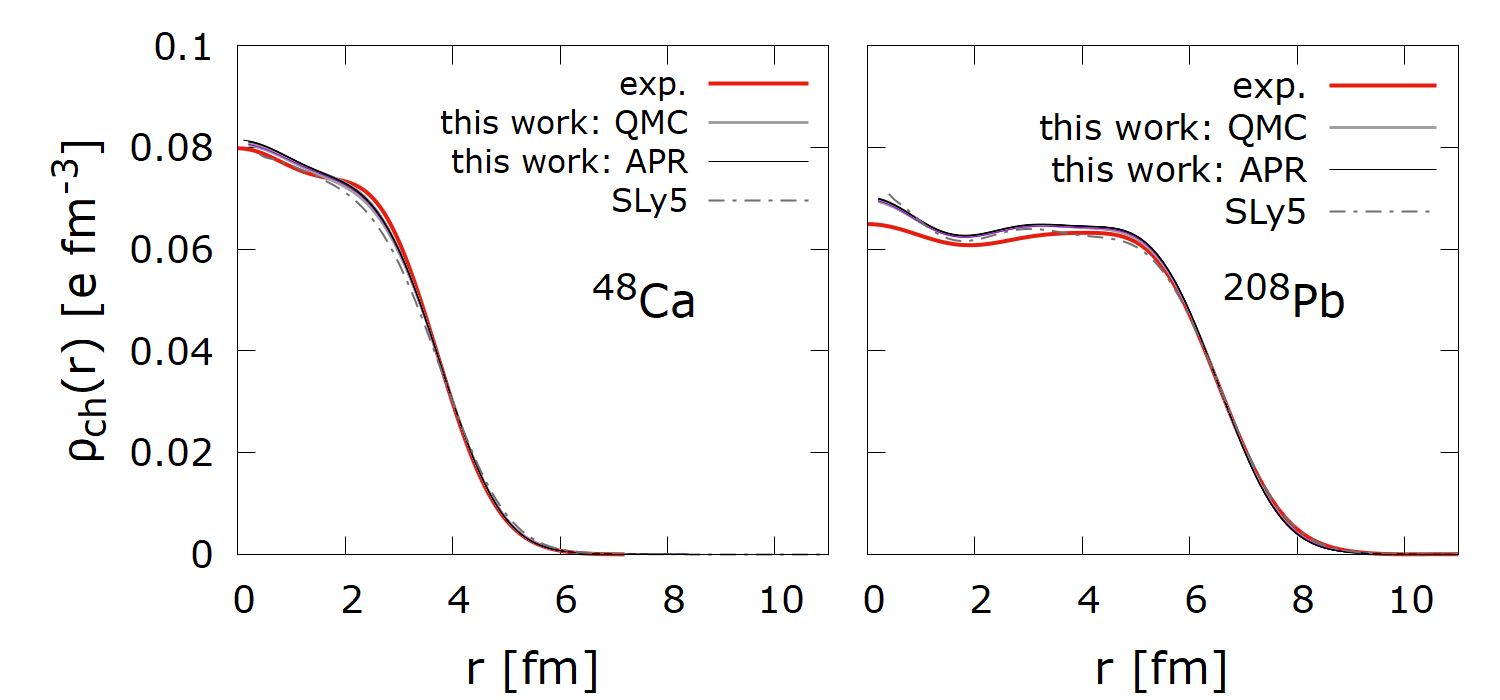}  
\caption{Charge density distributions obtained with a sample of APR- and QMC-based functionals and with the SLy5 functional compared with experimental results~\cite{ddd1987}. 
\label{fig:rhoch}}
\end{figure}

Charge radii of closed-shell nuclei have already been included among the optimization constraints through Eq.~(\ref{eq:adpd}). 
Figure \ref{fig:rhoch} compares the calculated charge density distributions with the experimental distributions and with the predictions of the standard Skyrme functional SLy5.  
The overall quality of the description is comparable to that obtained with SLy5, indicating that the low-density correction does not degrade the reproduction of measured charge distributions. No attempt has been made in the present work to re-optimize the bulk EDF parameters ($\rho_0,E_0,K_0,$, {\em etc.}), 
since the purpose of the low-density correction at present is to isolate the effect of the dilute-density sector while preserving the overall quality of the functional.

The neutron skin thickness, however, is a considerably more demanding observable than the charge radius because it is obtained as the difference between two much larger quantities, the neutron and proton radii.
A typical theoretical uncertainty in the charge radius, and therefore the proton radius $R_p$, is $0.02$~fm or larger. 
For most nuclei this corresponds to an error of less than $1\%$. 
The uncertainty in the neutron radius is expected to be at least comparable.
The neutron skin is therefore expected to carry an uncertainty of the same order. 
This already amounts to roughly $10-20\%$  for a typical neutron skin, 
illustrating the loss of relative precision when subtracting two much larger quantities.

Inspection of the point-nucleon density distributions reveals a reduction of their diffuseness when $V_d$ is introduced. 
A different diffuseness for the less constrained neutron distribution can in turn result in different $R_{np}$ values as discussed for example in~\cite{WVR2010}. 
The surface diffuseness can also affect the higher orders of the charge density distribution, which are not constrained as well as the mean square radius~\cite{CPP2026}.  

Theoretical uncertainties associated with the surface diffuseness are taken into account in the analyses of the parity-violating asymmetry measurements~\cite{HKM2014}. 
Exploring the contributions and interplay of $V_d$, $\wiv$, $L$, and other parameters would be of interest 
for interpreting future parity-violating measurements at MREX, which will further probe the surface regime~\cite{Sch2024}.

Arguably, one is not obligated to reproduce the PREX-II result within its large uncertainty~\cite{Sam2023}. 
Without the PREX-II constraint, tensions dissolve~\cite{XXL2020,ETL2021,ABD202X}. 
Even so, the present results 
reveal a meaningful degree of freedom worth addressing in nuclear energy density functional theory. \textcolor{black}{The low-density sector of the functional remains underconstrained unless a broader set of observables probing the nuclear surface is included, such as neutron skins and the fourth moment of the charge density distribution~\cite{RNG2020,NCL2021,CPP2026}.
} 

\section{Conclusion}

The CREX and PREX-II neutron-skin measurements of 
$^{48}$Ca and $^{208}$Pb are reproduced and reconciled with their dipole polarizabilities within nuclear energy density functional theory consistent with realistic neutron-star EoSs. 
This is achieved by relaxing the assumption, implicit in current EDF parameterizations, that the low-density sector follows as a smooth extrapolation of uniform, liquid-like saturated matter. 
By allowing controlled variations of the low-density regime relevant to the nuclear surface, 
the apparent CREX–PREX-II inconsistency is resolved at moderate values of the slope parameter $L$. 

More generally, the present results point to an underconstrained degree of freedom in current EDFs associated with dilute nuclear matter. Constraining this sector will require observables with enhanced sensitivity to the nuclear surface, such as neutron skins and higher moments of the charge density distribution. Future parity-violating electron-scattering experiments and other surface-sensitive measurements therefore have the potential to further constrain the low-density behavior of nuclear energy density functionals.

\vspace{2mm}

{\em Acknowledgements---}The author wishes to thank Bao-An Li and Joseph B. Natowitz for sharing their insights and suggestions. 
Thanks also go to the participants and organizers of the {\em IRL-NPA-FRIB Workshop on Dense Nuclear Matter Equation of State from Theory and Experiments}, MSU, East Lansing (2024) and the  {\em Fourth Workshop on Density Functional Theory: Fundamentals, Developments, and Applications}, RCNP, Osaka (2026) from which this work benefited. 
This work was supported by the Institute for Basic Science (IBS) through
the National Research Foundation (NRF) of Korea (2013M7A1A1075764) and by the NRF (TOPTIER, RS-2024-00436392).

\appendix 

\section{Basic formulas of the KIDS framework\label{Sec:kids}}

For completeness, the working expressions of the KIDS framework are reproduced, in standard notation, from Ref.~\cite{PaH2023}.
The physical motivation and the underlying formalism are discussed in \cite{kidsnm,kids_nuclei1} and reviewed in \cite{PaH2023}. 
Expressing the functional in Skyrme-type notation makes it straightforward to adapt existing Skyrme-based Hartree-Fock and RPA codes to the KIDS framework.
The convention $y_i\equiv t_ix_i$ is used. 

The energy per particle, originally formulated for homogeneous matter as an expansion in powers of the Fermi momentum, 
can be written in the convenient form of an extended Skyrme functional,
\begin{eqnarray}
\epa &=& \EDF / \rho  = \tpa + \frc{3}{8}t_0\rho 
-\frc{1}{8} (t_0+ 2y_0) \rho_3^2/\rho 
\nonumber \\ && 
+ \sum_{i=1}^{\imaxx} [\frc{1}{16}t_{3i}\rho^{1+i/3} 
- \frc{1}{48}(t_{3i}+2y_{3i})\rho^{-1+i/3}\rho_3^2 ]  \nonumber 
\\ 
 & & 
+\frc{1}{16}(3t_1 + 5t_2 + 4y_2) \rho\tau - \frc{1}{16} (t_1 + 2y_1 - t_2 - 2y_2)\rho_3\tau_3
\nonumber \\ 
 & &  
+\frc{1}{64}(-9t_1 + 5t_2 + 4y_2) \nabla^2\rho 
\nonumber \\ && + \frc{1}{64} (3t_1 + 6y_1 + t_2 + 2y_2)(\rho_3/\rho )\nabla^2\rho_3 
\nonumber \\ 
 & & 
 -\frc{1}{2}W_0{\bf \nabla \cdot J}  -\frc{1}{2}W_1(\rho_3/\rho ) {\bf \nabla \cdot J}_3 , 
\label{Eq:KIDSEDF}
\end{eqnarray}
where $\tpa = \frac{3}{5}\frac{\hbar^2}{2m}(3\pi^2)^{2/3} (\rho_p^{5/3}+\rho_n^{5/3})/\rho$ 
denotes the energy per particle of the free Fermi gas while $\rho_3 \equiv \rho_n-\rho_p$. 
The energy per particle $\epa$  can be generated at the Hartree-Fock level from an effective interaction 
\begin{eqnarray}
V (1,2) &=& (t_0 + y_0 \hat{P})\! +\!\! 
\mbox{$\frac{1}{6}$}\sum_{i=1}^{\imaxx} (t_{3i}+y_{3i}\hat{P})\rho^{i/3}({\bf R}_{12}) \delta ({\bf r}_{12}) 
\nonumber \\
&& + \frc{1}{2}(t_1+y_1\hat{P})[{\bf p}_{12}^2\delta({\bf r}_{12}) +  \delta({\bf r}_{12}){\bf p}_{12}^2] 
\nonumber  \\ && 
+ (t_2+y_2\hat{P}){\bf p}_{12}\cdot \delta({\bf r}_{12}){\bf p}_{12} 
\end{eqnarray} 
plus spin-orbit terms. 
In this work, $\imaxx=3$ is used. 

Setting for brevity $h_k\equiv \frac{3}{5}(3\pi^2)^{2/3}$, the energy per particle of symmetric nuclear matter is given by 
\beq
E(\rho,0) = (\frc{\hbar^2}{2m}+\ceff \rho ) {h_k}{2^{-2/3}} \rho^{2/3}  + \sum_{i=0}^{\imaxx} \alpha_i \rho^{1+i/3} 
\label{eq:e0kids} 
\eeq
while the symmetry energy  is given by 
\beq 
S(\rho ) = \sigma (\frc{\hbar^2}{2m} + \ceff\rho )h_k\rho^{2/3} + \varphi\deff h_k\rho^{5/3} +\sum_{i=0}^{\imaxx}\beta_i\rho^{1+i/3} ,
\label{eq:srhokids} 
\eeq
where $\sigma =\frac{5}{9}\frac{1}{2^{2/3}}$ 
and $\varphi = \frac{5}{3}\frac{1}{2^{2/3}}$. 
The kinetic parameters $\ceff,\deff$ determine the isoscalar and isovector in-medium effective masses. At saturation density one has  
\[ 
\mu_s \equiv m^{\ast}_s/m = [ 1 + \frc{2m}{\hbar^2}\rho_0\ceff ]^{-1}  ,
\]
\[
\mu_v \equiv m^{\ast}_v/m =  [ 1 + \frc{2m}{\hbar^2}\rho_0(\ceff - \deff ) ]^{-1}.
\] 
Finally, the contribution of the gradient terms to the energy per particle is denoted as  
\begin{equation} 
\epa_{\mathrm{grad}} = C_{12} \nabla^2\rho  + D_{12} (\rho_3/\rho )\nabla^2\rho_3 
,
\label{Eq:cd12} 
\end{equation} 
where $C_{12}=\frc{1}{64}(-9t_1 + 5t_2 + 4y_2)$, $D_{12}=\frc{1}{64} (3t_1 + 6y_1 + t_2 + 2y_2)$. 

For the present work, the key step is to transform any given set of bulk EoS and EDF parameters 
$(E_0,\rho_0,K_0)$, $(J,L,K_{\mathrm{sym}},Q_{\mathrm{sym}})$, $(\ceff,\deff,C_{12},D_{12} )$ into a functional (\ref{Eq:KIDSEDF}). 
The spin-orbit parameters decouple from the remaining coefficients and are treated independently. 
The momentum-dependence coefficients, $t_{1,2},y_{1,2}$, are given by  
\beq 
\left( 
\begin{array}{c}  
t_1 \\ y_1 \\ t_2 \\ y_2 
\end{array} 
\right) 
= 
\frac{2}{3}
\left( 
\begin{array}{rrrr}  
2 & 0 & -8 & 0  \\  
-1 & -3 & 4 & 12  \\ 
6 & -12 & 8 & -16  \\ 
-3 & 15 & -4 & 20  
\end{array} 
\right) 
\left( 
\begin{array}{c}  
\ceff \\ \deff \\ C_{12} \\ D_{12}  
\end{array} 
\right) .
\eeq
For the $\alpha_i$, $\beta_i$ expansion coefficients of Eqs.~(\ref{eq:e0kids}), (\ref{eq:srhokids})  one solves 
\begin{widetext} 
\beq 
\left(
\begin{array}{l} 
\alpha_0\rho_0 \\ \alpha_1\rho_0^{4/3} \\ \alpha_2 \rho_0^{5/3} 
\end{array} \right) 
= \left( 
\begin{array}{rrc}  
10   &   -3   &   1/2  \\
 -15  &  5  &  -1  \\
 6  &  -2  &  1/2  
\end{array} 
\right) 
\left(\begin{array}{r} 
E_0 - [\frc{\hbar^2}{2m} + \ceff\rho_0]h_k(\rho_0/2)^{2/3} \\
 - [2\frc{\hbar^2}{2m} + 5\ceff\rho_0]h_k(\rho_0/2)^{2/3} \\ 
K_0 + [2\frc{\hbar^2}{2m} -10 \ceff\rho_0]h_k(\rho_0/2)^{2/3}
\end{array} 
\right) ,
\label{Eq:SNM3}
\eeq 
\beq 
\left(
\begin{array}{l} 
\beta_0\rho_0 \\ \beta_1\rho_0^{4/3} \\ \beta_2 \rho_0^{5/3} \\ \beta_3\rho_0^2
\end{array} \right) 
= \left( 
\begin{array}{rccr}  
20   &   -19/3   &   1 & -1/6  \\
 -45  &  15  &  -5/2 & 1/2   \\
 36  &  -12  &  2  & -1/2 \\
-10 & 10/3 & -1/2 & 1/6  
\end{array} 
\right) 
\left(\begin{array}{r} 
J - [\sigma(\frc{\hbar^2}{2m} + \ceff\rho_0) + \varphi\deff\rho_0 ]h_k\rho_0^{2/3} \\
L - [\sigma(2\frc{\hbar^2}{2m} + 5\ceff\rho_0) + 5\varphi\deff\rho_0 ]h_k\rho_0^{2/3} \\ 
\Ksym + [\sigma (2\frc{\hbar^2}{2m} -10 \ceff\rho_0) -10\varphi\deff\rho_0 ]h_k\rho_0^{2/3}\\ 
\Qsym + [\sigma(-8\frc{\hbar^2}{2m} +10 \ceff\rho_0) + 10\varphi\deff \rho_0]h_k\rho_0^{2/3}
\end{array} 
\right)
.
\label{Eq:SE} 
\eeq
\end{widetext} 
Following \cite{kidsnm,kids_nuclei2},  $\alpha_3$ is set to zero, but the expressions can be extended to any order \cite{PaH2023}. 
Finally, the standard Skyrme-like parameters are obtained as  
\beq 
t_0 = \frc{8}{3}\alpha_0 \, ; \, t_{31}=16\alpha_1 \, ; \, t_{32} = 16 \alpha_{2} \, ; \, t_{33}=16 \alpha_{3}= 0
\eeq 
and 
\beq 
y_0 = -\frc{1}{2}t_0 - 4\beta_0  \quad ; \quad y_{3i}= -\frc{1}{2} t_{3i} - 24\beta_{i} \quad (i=1,2,3) . 
\eeq

\section{Treatment of dilute matter \label{Sec:Vd}} 

In principle, one could construct two different energy density functionals for saturated and dilute nuclear matter
(for example according to (\ref{Eq:KIDSEDF})), $\epa_{\mathrm{bulk}}$ 
and $\epa_{\mathrm{dil}}$, 
and interpolate smoothly between them below a matching density $\rho_c$, where $\rho_c$ is an unspecified fraction of the saturation density.
To avoid introducing a large number of additional parameters, only the difference between the two is parameterized here instead,
\[ 
\mathcal{E}_d = \epa_{\mathrm{dil}}-\epa_{\mathrm{bulk}}= \epa_{\mathrm{dil}}-\epa \propto \rho^{a_d}e^{-b\rho^2} ,
\]
to be added to $\epa$ of Eq.~(\ref{Eq:KIDSEDF}). 
This is equivalently implemented by adding to $V(1,2)$ the term $V_d$ of Eq.~(\ref{eq:Vd}). 
The contribution to the energy density is 
\[ 
\mathcal{H}_d = \left( \frac{t_d}{16} + \frac{t_d+2y_d}{48}\frac{\rho_3^2}{\rho^2}\right) \rho^{2+a_d}e^{-b_d\rho^2} . 
\]
This is the contribution used throughout the Hartree-Fock and RPA calculations.
The exponential factor ensures that the correction becomes negligible well below saturation density, so that the properties of homogeneous matter around and above saturation remain unaffected.

Hartree-Fock and RPA calculations require the first and second derivatives of the energy density 
with respect to the proton and neutron densities in order to construct the single-particle potentials, 
rearrangement terms, and residual interaction.
The derivations are straightforward and the results reduce, for $b_d=0$, 
to the familiar expressions obtained from the standard density-dependent $(t_3,y_3)$ terms of the Skyrme functional. 
As for the standard density-dependent terms, the contribution of the present correction is purely additive to the energy density, the energy per particle, the single-particle potential, and the residual interaction.

\input{paper.bbl}

\end{document}

%% file: paper.bbl
%